\documentclass[preprint,showpacs,preprintnumbers,amsmath,amssymb,nofootinbib]{revtex4-1}

\usepackage{graphicx}
\usepackage{dcolumn}
\usepackage{amssymb}
\usepackage{mathrsfs}
\usepackage{amsmath}
\usepackage{epsfig}
\usepackage{hhline}
\usepackage{color}
\usepackage{multirow}
\usepackage[T1]{fontenc}
\usepackage{verbatim}
\usepackage{slashed}
\usepackage{hyperref}

\begin{document}

\title{Probing a new decay of vector-like top partner mediated by heavy Majorana neutrino via single production}
\author{Hang Zhou}
\author{Ning Liu}
\email[Corresponding author: ]{liuning@njnu.edu.cn}
\affiliation{Physics Department and Institute of Theoretical Physics, Nanjing Normal University, Nanjing, 210023, China}

\begin{abstract}
Models beyond the Standard Model have been proposed to simultaneously solve the problems of naturalness and neutrino mass, in which heavy Majorana neutrinos and vector-like top partners are usually predicted. A new decay channel of the top partner mediated by the heavy Majorana neutrino can thus appear: $T\to b\,W^{+}\to b\,\ell^{+}\ell^{+}q\bar{q'}$. We study in this paper the observability of this decay process through single production of the top partner at the 14 TeV LHC: $pp\to T/\bar{T}$+jets$\to b/\bar{b}+\mu^{\pm}\mu^{\pm}$+jets. $2\sigma$ exclusion bounds on the top partner mass and mixing parameters are given by Monte-Carlo simulation, which surpass those from the search through VLT pair production in the mass range of $m_{T}>1.3$ TeV.
\end{abstract}
\maketitle

\section{Introduction}

The discovery of the 125\,GeV Higgs boson at the LHC in 2012~\cite{Aad:2012tfa,Chatrchyan:2012xdj} marks a great success of the Standard Model (SM) and deepens our understanding of the electroweak symmetry breaking. With a mass at the electroweak scale ($\sim10^{2}$\,GeV), the observed Higgs boson causes the so-called naturalness problem: the Higgs mass receives loop corrections from heavy particles like the SM top quark, which can lead the Higgs mass to Planck scale unless new physics is present to cancel out the quadratical divergence. The naturalness problem motivates a variety of new models beyond SM (BSM), such as the composite Higgs~\cite{Dugan:1984hq,Kaplan:1991dc,Contino:2006qr,Contino:2006nn} and the little Higgs models~\cite{Perelstein:2003wd,Matsedonskyi:2012ym}, through the introduction of a spontaneously broken global symmetry that leads the Higgs boson to be a pseudo Goldstone boson. Vector-like top partners (VLT) are usually present in these models and play an important role in the cancelling of the quadratical divergence in the Higgs mass from the SM top loop. Therefore, VLTs have been widely studied and searched for at hadron colliders through both single and pair production, with subsequent decays into a SM quark and a gauge boson or Higgs boson~\cite{delAguila:2000aa,delAguila:2000rc,AguilarSaavedra:2009es,Han:2014qia,Liu:2015kmo,Yang:2018oek,Liu:2019jgp,Chala:2017xgc,Zhou:2019alr}. ATLAS and CMS collaborations at the LHC excluded VLT with mass lower than $740\sim1370$\,GeV, depending on its SU(2) representation and different branching ratios assumed~\cite{ATLASbounds,CMSbounds}.

On the other hand, observation of neutrino oscillation from atmospheric, solar, reactor and accelerator experiments indicates neutrinos of different flavors are mixed and massive around sub-eV scale~\cite{Tanabashi:2018oca}. Seesaw mechanism~\cite{minkowski1977,yanagida1979,ms1980,sv1980,Weinberg:1979sa,mw1980,cl1980,lsw1981,ms1981,flhj1989,ma1998}, among various schemes to include neutrino mass into the SM, is one of the most popular.  By introducing three right-handed (RH) neutrinos, the Type-I seesaw~\cite{minkowski1977,yanagida1979,ms1980,sv1980} can naturally generate sub-eV Majorana neutrino masses if the RH Majorana masses are about $\sim10^{14}$\,GeV while the Dirac masses remain at the electroweak scale. Seesaw mechanism links the origin of neutrino mass with the observed baryon asymmetry through leptogenesis~\cite{fy1986,krs1985,lpy1986,luty1992,mz1992,fps1995,crv1996,pilaftsis1997,ms1998,hs2004,Gu:2019nhb}, some variations and extensions of which can also accommodate dark matter particles~\cite{ma2006,ma2015,Zhou:2016jyp,Gu:2019ogb}. Majorana neutrinos are key features of the seesaw models, the generation of which always goes with lepton number violation (LNV) by 2 units. Thus the searches for neutrinoless double beta decay~\cite{Furry:1939qr,Doi:1985dx} and other LNV processes~\cite{Ng:1978ij,Abad:1984gh,Littenberg:1991rd,Dib:2000wm,Ali:2001gsa,Barbero:2002wm} have been performed to test the Majorana nature of neutrinos. Depending on whether the mediator neutrino is light or heavy compared with the LNV scale, the LNV processes are suppressed either by a factor of $m^{2}_{\nu}/m^{2}_{W}$ due to the light neutrino mass $m_{\nu}$, or by a factor of $|V_{\alpha m}V_{\beta m}|^{2}$ due to their small mixings~\cite{Atre:2009rg}. However, if the heavy Majorana neutrino mass can be kinematically accessible (below TeV) as in some low-scale Type-I seesaw scenarios~\cite{Asaka:2005an,Asaka:2005pn,Asaka:2006nq,Asaka:2006ek,Xing:2009in,Adhikari:2010yt,Ibarra:2010xw,Boucenna:2014zba,Zhou:2017lrt,Gu:2018kmv}, the LNV processes can be substantially enhanced by resonant production of the heavy neutrinos, which may be directly searched for at colliders ~\cite{deGouvea:2006gz,deGouvea:2007hks,Atre:2009rg,Kersten:2007vk,Bajc:2007zf,He:2009ua,Han:2006ip,Ibarra:2011xn,Dev:2013wba,Deppisch:2015qwa,Das:2018hph,Liu:2019qfa}. LEP experiments have put an upper limit on the mixing $|V_{\mu N}|^{2}<\mathcal{O}(10^{-5})$ for heavy neutrino mass of 80\,GeV$\sim$205\,GeV~\cite{Abreu:1996pa} and CMS has given a similar bound of $|V_{eN,\mu N}|^{2}<\mathcal{O}(10^{-5})$ for a broader mass range 20\,GeV$\sim$1600\,GeV~\cite{Sirunyan:2018mtv,Sirunyan:2018xiv}. The much more stringent bound on $|V_{eN}|^{2}$ ($10^{-8}\sim10^{-7}$) was given by GERDA experiments~\cite{Agostini:2018tnm}.

Models have been proposed to solve the above two BSM issues simultaneously by incorporating neutrino mass into scenarios with VLT. For example, LNV interaction between triplet scalar and doublet lepton can be included within the Littlest Higgs scenario~\cite{Han:2005nk}. Other examples include Little Higgs models~\cite{delAguila:2019mvp,Dey:2008dk,deAlmeida:2007khx,Li:2011ao,Hektor:2007uu,Goyal:2006yn,Abada:2005rt,Goyal:2005it,delAguila:2005yi,Lee:2005mba,Chang:2003vs}, Composite Higgs models~\cite{Coito:2019wte,Shindou:2017bem,Smetana:2013hm,delAguila:2010vg,Lee:2005kd}, Higgs Inflation models~\cite{He:2014ora}, Top Seesaw models~\cite{He:1999vp,He:2001fz,Wang:2013jwa}, etc~\cite{Du:2012vh,Abe:2012fb}. VLT and heavy Majorana neutrinos are what these BSM models have in common and hence a new decay channel of VLT will be present through a mediating heavy Majorana neutrino. As mentioned above, VLTs and heavy Majorana neutrinos can both be searched for at the LHC, we thus propose a model-independent search strategy for the new decay channel of VLT in a scenario that includes three RH Majorana neutrinos and a singlet top partner $T$. As the mass of VLT increases, the cross section of its single production will surpass pair production at the LHC, as a result of the collinear enhancement of the light quark emitting a $W$ boson~\cite{Willenbrock:1986cr}. Besides, the single production of VLT also has a unique event topology that can be used to suppress the SM backgrounds. Therefore, we focus on the VLT single production as a complementary study of the search by pair production~\cite{Zhou:2020ovl}. We will demonstrate in the rest of the paper that with GeV-scale Majorana neutrinos, the new decay channel of VLT can be probed at the LHC by searching for final same-sign dileptons~\cite{Cao:2011ew}. In the next section we will introduce relative effective Lagrangian of the present scenario. Section III is our analysis by Monte-Carlo simulation of the search at the 14 TeV LHC and exclusion limits will be given on the VLT couplings and Majorana neutrino mixings. Section IV is our conclusion.

\section{The new decay mode of VLT and the relevant Lagrangian}

\begin{figure}[b]
\centering
\begin{minipage}{0.49\linewidth}
  \centerline{\includegraphics[scale=0.305]{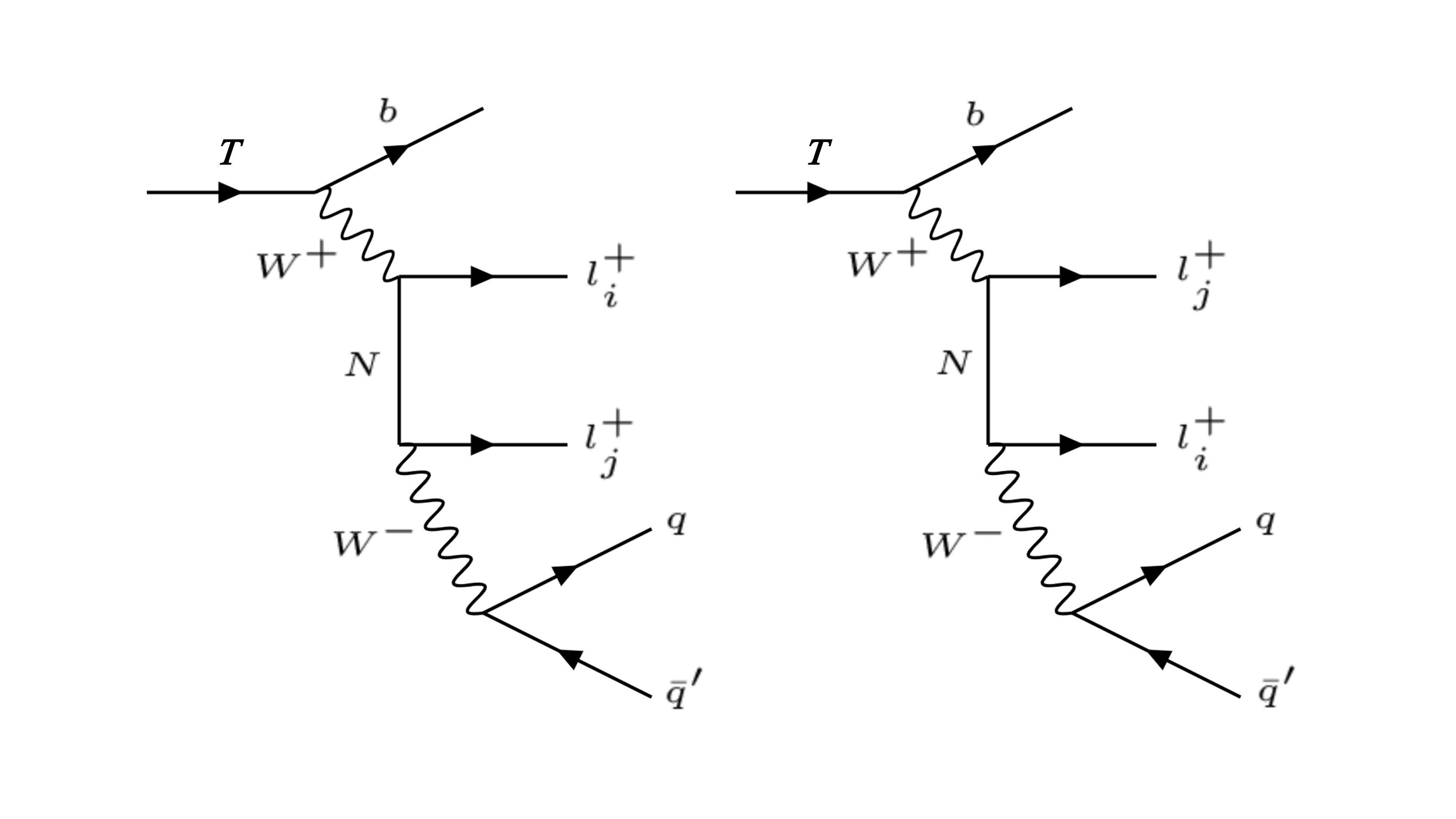}}
  \centerline{(a)}
\end{minipage}
\hfill
\begin{minipage}{0.49\linewidth}
  \centerline{\includegraphics[scale=0.64]{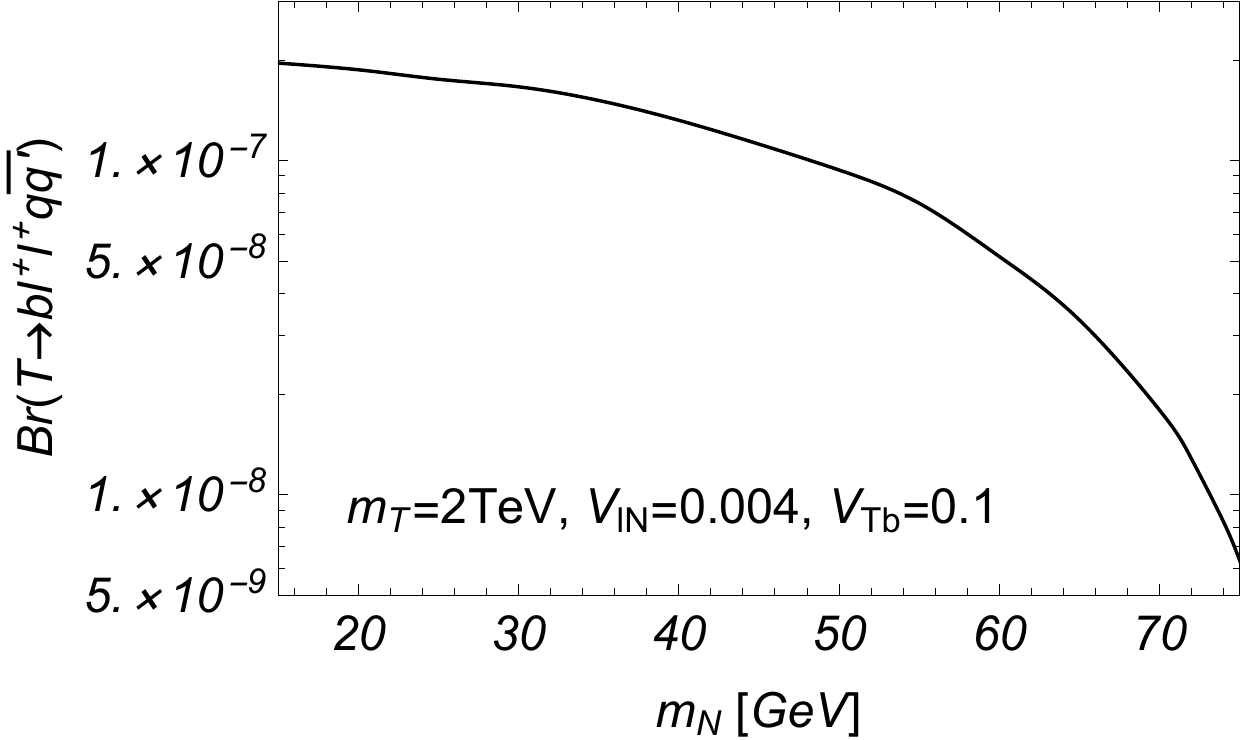}}
  \centerline{ }
  \centerline{(b)}
\end{minipage}
\caption{Feynman diagrams and branching ratio of the top partner decay $T\to b\,\ell^{+}\ell^{+}q\bar{q'}$ including t- and u-channels.}
\label{fig:fdandbr}
\end{figure}

As a phenomenological investigation and without losing general features, we parameterize the low-scale Type-I seesaw by a single right-handed Majorana neutrino mass $m_{N}$ and a mixing parameter between the light and heavy neutrinos $V_{\ell N}$. Introduction of interactions between VLT and gauge bosons will lead to a new decay mode of $T$ through mediating heavy Majorana neutrino, ending up with a pair of same-sign leptons (\figurename~\ref{fig:fdandbr}(a)): $T\to b\,W^{+}\to b\,\ell^{+}\ell^{+}q\bar{q'}\,$. We will show in the next section that the same-sign dilepton in the final state can serve as a special signature at the LHC to search for this new decay mode. The effective interactions relevant to the VLT decay process are
\begin{align}
\mathcal{L}={}&-\frac{g}{2\sqrt{2}}W^{+}_{\mu}\left[V_{\ell N}\ell\gamma^{\mu}(1-\gamma_{5})N^{c}+V_{Tb}\bar{T}\gamma^{\mu}(1-\gamma_{5})b\right]+\textrm{H.c.}\,,
\label{Ltotal}
\end{align}
in which $V_{\ell N}$ is mixing parameters between the light-flavor and heavy Majorana neutrinos, $N$ refers to three heavy Majorana neutrinos, $\ell$ here marks charged leptons of three flavors: $e$, $\mu$, and $\tau$. $V_{Tb}$ is coupling of the top partner $T$ with $W$ boson.

Depending on mass of the heavy Majorana neutrino $m_{N}$, branching ratio of the above VLT decay is presented in \figurename~\ref{fig:fdandbr}(b), assuming $\text{Br}(T\to bW^{+})=50\%$. We also assume in the calculation $m_{T}=2$\,TeV, $V_{\ell N}=0.004$ and $V_{Tb}=0.1$ that are not excluded by current experiments (But note that $V_{\mu N}=0.004$ survives from $2\sigma$ bounds~\cite{Abreu:1996pa} in the mass range of \figurename~\ref{fig:fdandbr}(b) while $V_{eN}=0.004$ does not~\cite{Agostini:2018tnm}). As $m_{N}$ grows larger than $m_{W}$, the rare decay will be enhanced as the result of on-shell production of $W$ boson from $N$ decay, but the enhancement is not that large and the rare decay branching ratio ($\sim10^{-8}$) is still lower than that in the light mass range ($m_{N}\lesssim m_{W}$). Therefore in the next section we focus on this mass range of the heavy Majorana neutrino and study the search at the LHC for the VLT new decay mode. Note that the rare decay mode of the SM top quark $t\to b\,\ell^{+}\ell^{+}jj$, comparing with the one of the top partner $T$, can actually be a more frequent possibility and be used to study the light-heavy neutrino mixings~\cite{Liu:2019qfa}. While in the present scenario that accommodates neutrino masses and naturalness, the search for $T$ decay $T\to b\,\ell^{+}\ell^{+}jj$ can provide information for both the seesaw mechanism and the top partner simultaneously. It should also be noted that large neutrino mixings can be inconsistent with small neutrino masses and a Majorana singlet of $m_{N}\lesssim m_{W}$, but this can be resolved by introducing two bi-spinors per family~\cite{Hernandez-Tome:2019lkb}. Fine-tuning should also be required to cancel out the radiative corrections, but for the above mass range below electroweak scale, the lepton number violating signature can be observable at the LHC without fine-tuning as the result of destructive interference between contributions from different neutrinos~\cite{Drewes:2019byd}.

\section{Search for the new decay at the LHC}

The SU(2) singlet VLT can be produced singly through proton-proton collision at the LHC via electroweak interactions: $pp\to Tq\,/T\bar{q}\,/TW$, among which the $W$-exchange production ($qb\to Tq$) has the largest cross section. The singly produced VLT can then go through the new decay mode $T\to b\,\ell^{+}\ell^{+}jj$. If the $W$ boson accompanied with VLT decays hadronically, we will have the signal of a same-sign dilepton and multijets including a b-tagged one: $pp\to Tq\,/T\bar{q}\,/TW^{-}\to b+\ell^{+}\ell^{+}+\text{multijets}$.
Given the fact that $e$-flavor mixing with heavy neutrino $|V_{eN}|^{2}$ has been strictly bounded below $\sim10^{-8}$ in the mass range from GeV to $10^{2}$\,GeV by GERDA experiments~\cite{Agostini:2018tnm} and that high efficiency and accuracy of $\tau$-tagging are necessary for limiting $\tau$-flavor mixing, which are beyond the ability of current collider simulation, we focus on the dimuon channel in the mass range of heavy Majorana neutrino ($m_{N}\lesssim m_{W}$) that is kinematically accessible at the LHC for its resonant production. The contribution from CP-conjugation of the above process is also included in the simulation below. Therefore our signal process can be expressed as
\begin{align}
pp\to T/\bar{T}+\text{jets}\to b/\bar{b}+\mu^{\pm}\mu^{\pm}+\text{jets}.
\label{signal}
\end{align}
in which we consider mainly the $W$-exchange single production. Besides, a diagonalized mixing matrix between light-flavor and heavy neutrinos $V_{\ell N}$ is adopted and hence for the dimuon channel in our case, the mediated heavy Majorana neutrino is $N_{2}$ that couples exclusively to $\mu$-flavor.

As for the SM backgrounds for the signal consisting of a same-sign dimuon plus multijets, the major ones come from events with fake leptons (such as top pair production $t\bar{t}$ and single production $t/\bar{t}$+jets) and prompt multileptons (such as $t\bar{t}\,W^{\pm}$ and $W^{\pm}W^{\pm}$+jets). Therefore the following four kinds of processes are considered as backgrounds
\begin{align}
pp\to t\bar{t}\,,t/\bar{t}+\text{jets}\,,t\bar{t}\,W^{\pm}\,,W^{\pm}W^{\pm}+\text{jets}\,.
\label{bkg}
\end{align}
We did not include events with opposite-sign dimuons, which may also contribute to the background if one of the dimuon's charge is mismeasured, as the mismeasurement rate of muon charge is generally low. Note that the top decay mediated by the heavy Majorana neutrinos:
\begin{align}
t/\bar{t}\to b/\bar{b}+\ell^{\pm}\ell^{\pm}+\text{jets}\,,
\end{align}
will also be present in our scenario and contribute in the backgrounds $t\bar{t}$, $t\bar{t}W^{\pm}$ and $t/\bar{t}+$jets. These events are included in our simulation for the backgrounds. Monte-Carlo simulations are then performed for the signal Eq.\eqref{signal} and backgrounds Eq.\eqref{bkg} at the 14\,TeV LHC, with the benchmark point as
\begin{align}
m_{N}=50\,\text{GeV},\quad V_{\mu N}=1.0\,,\quad m_{T}=2\,\text{TeV},\quad V_{Tb}=0.1\,,
\end{align}
in which $m_{N}$ stands for the mass of $N_{2}$ for simplicity, while for $N_{1}$ ($N_{3}$) that couples solely to $e$ ($\tau$), we assume a kinematically inaccessible mass 300\,GeV (1\,TeV). Signal and background events are generated at parton level using \textsc{MadGraph5\_aMC@NLO}~\cite{Alwall:2014hca} (version 2.6.7) with the NN23LO1 PDF~\cite{Ball:2012cx}, and then by \textsc{checkmate2} (version 2.0.26)~\cite{Dercks:2016npn}, go through parton showering and hadronization with \textsc{pythia-8.2}\,\cite{pythia} as well as detector simulation with tuned \textsc{delphes-3.4.1}~\cite{delphes}. Jet-clustering is done using \textsc{fastjet}~\cite{fastjet} with anti-$k_t$ algorithm~\cite{anti-kt}. B-tagging efficiency is assumed to be $70\%$ with MV2c20 algorithm~\cite{mv2c20} in the simulation. To account for contributions from higher order QCD corrections, the leading-order cross sections of $t\bar{t}$ and $t\bar{t}\,W^\pm$ are normalized to NNLO and NLO, respectively~\cite{Czakon:2011xx,Frixione:2015zaa}.

\begin{figure}[t]
\centering
\begin{minipage}{0.32\linewidth}
  \centerline{\includegraphics[scale=0.34]{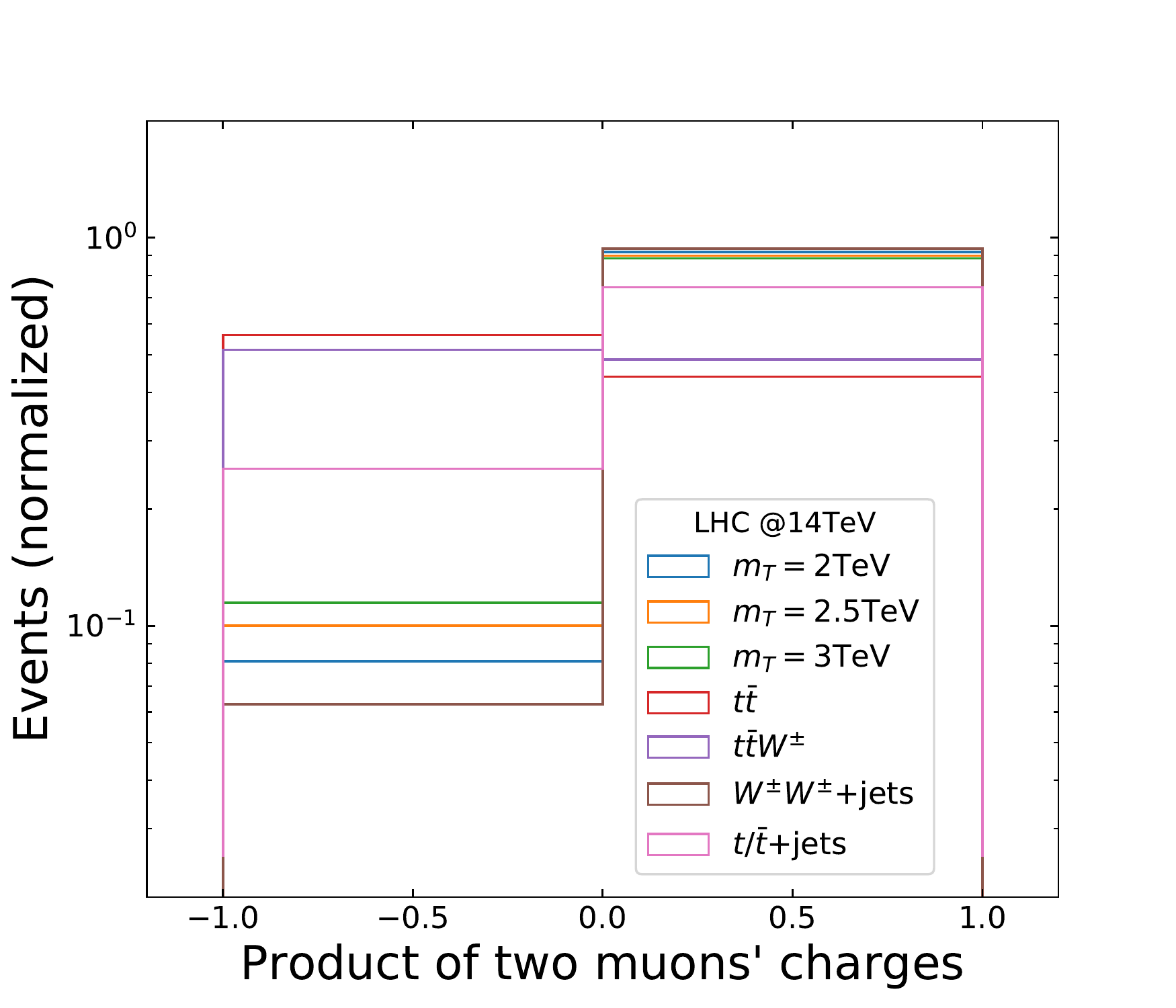}}
  \centerline{(a)}
  \label{fig:cc}
\end{minipage}
\hfill
\begin{minipage}{0.32\linewidth}
  \centerline{\includegraphics[scale=0.34]{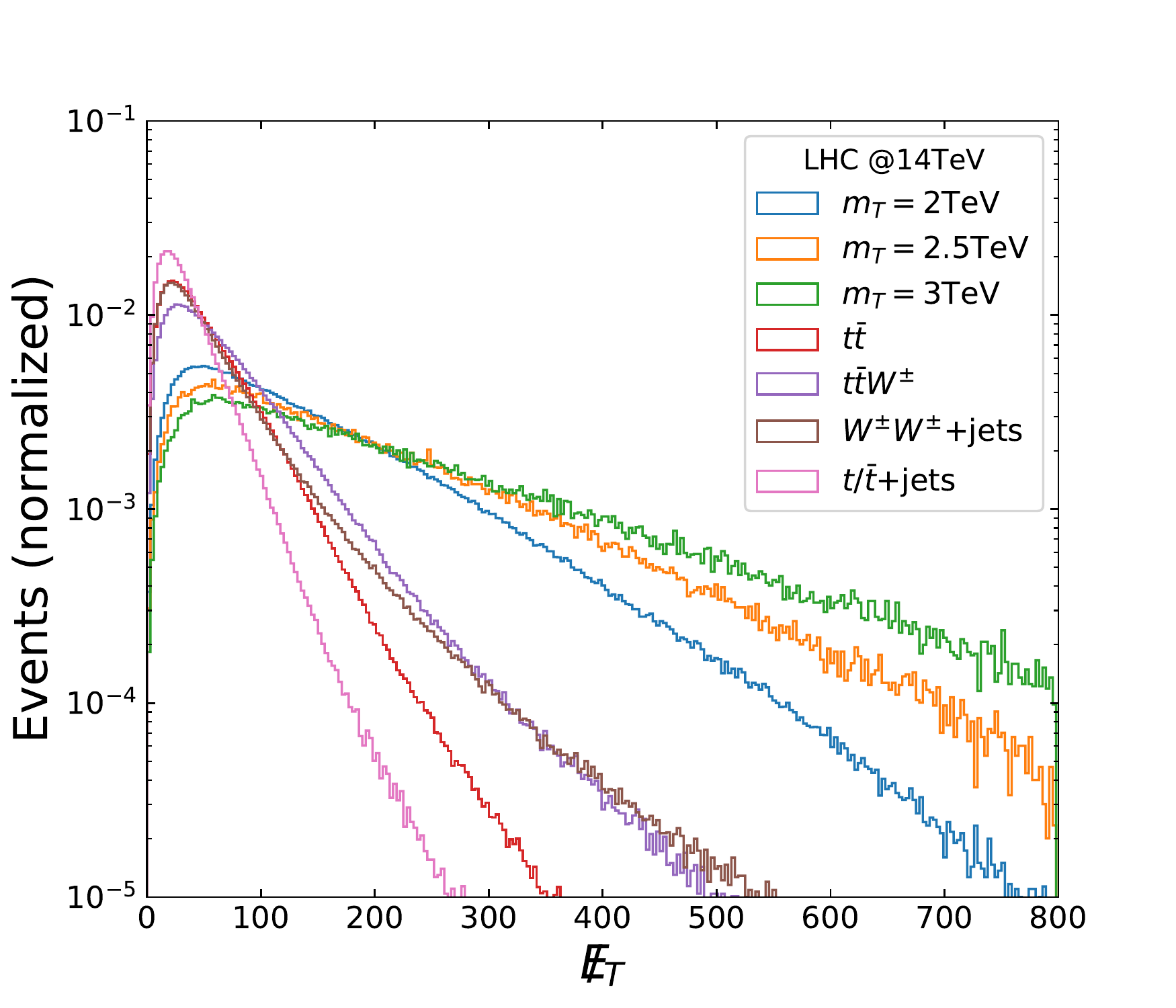}}
  \centerline{(b)}
\end{minipage}
\hfill
\begin{minipage}{0.32\linewidth}
  \centerline{\includegraphics[scale=0.34]{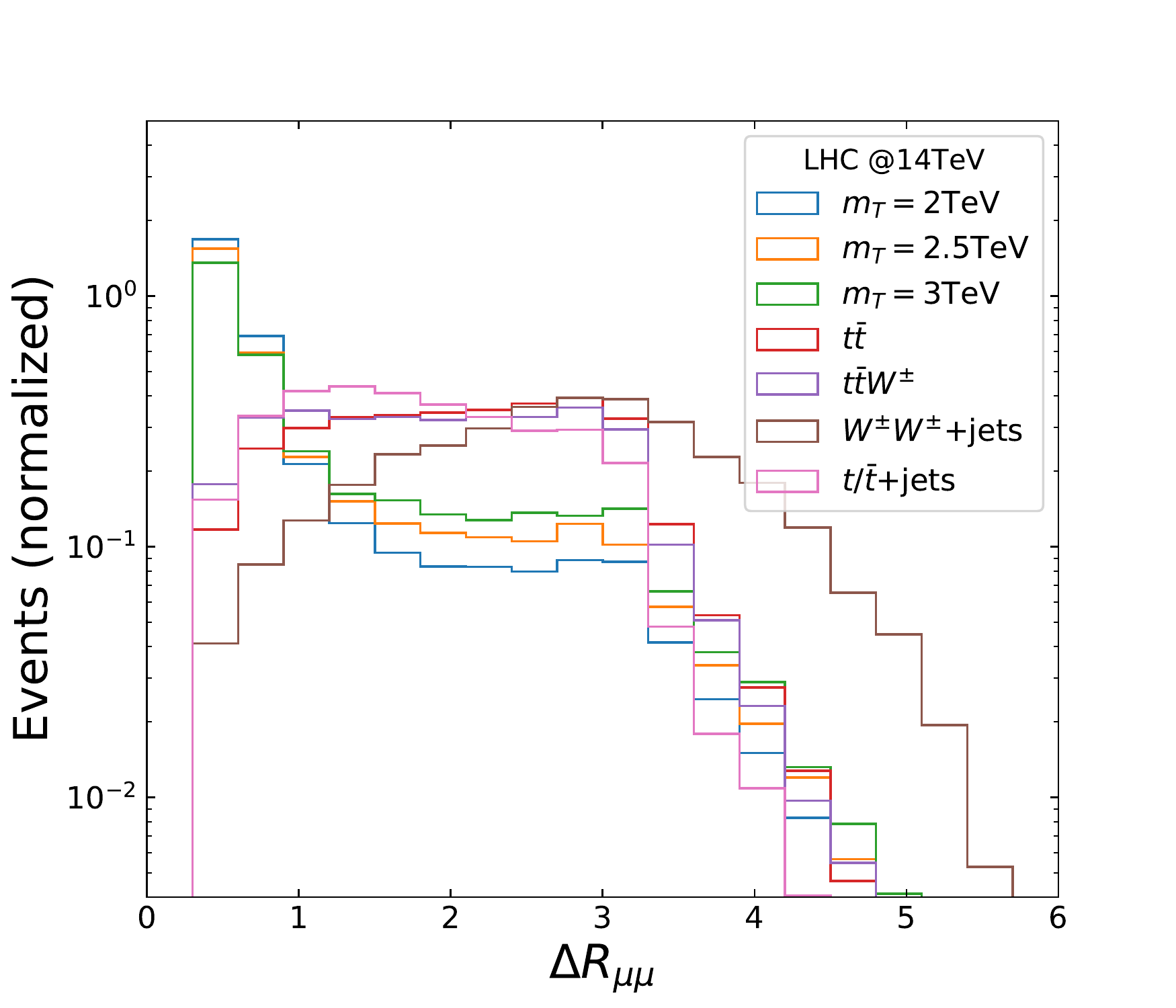}}
  \centerline{(c)}
\end{minipage}
\\
\begin{minipage}{0.32\linewidth}
  \centerline{\includegraphics[scale=0.34]{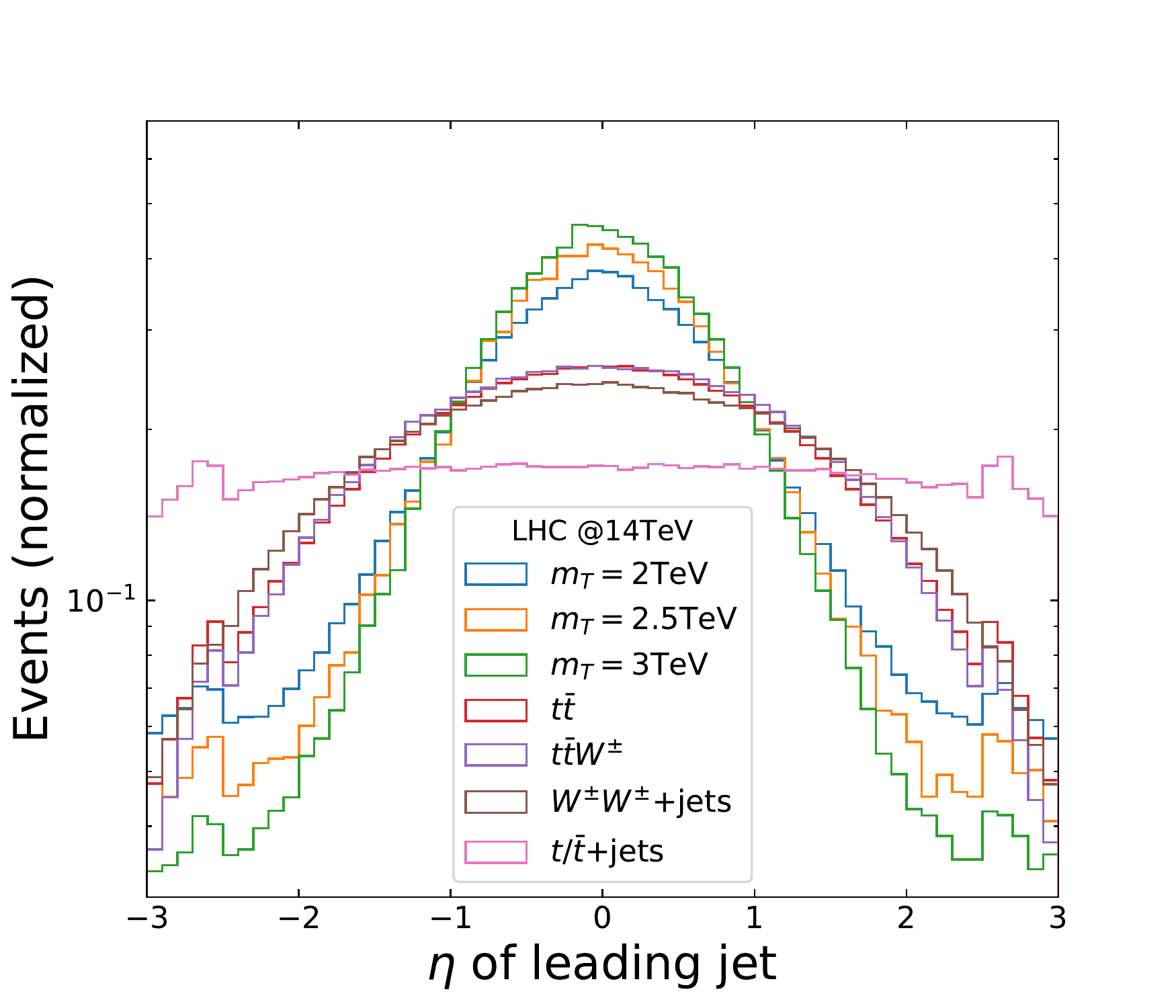}}
  \centerline{(d)}
\end{minipage}
\hfill
\begin{minipage}{0.32\linewidth}
  \centerline{\includegraphics[scale=0.34]{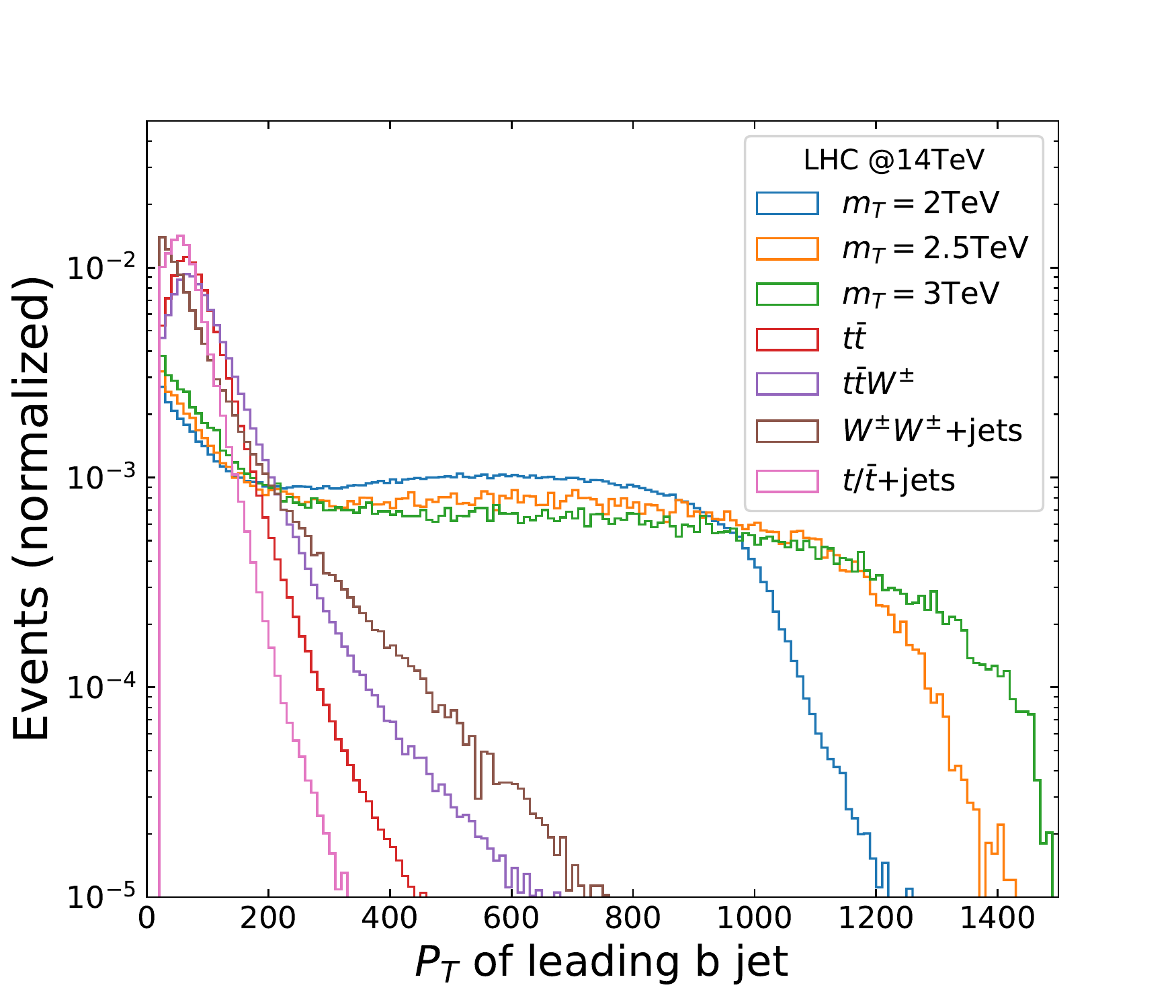}}
  \centerline{(d)}
\end{minipage}
\hfill
\begin{minipage}{0.32\linewidth}
  \centerline{\includegraphics[scale=0.34]{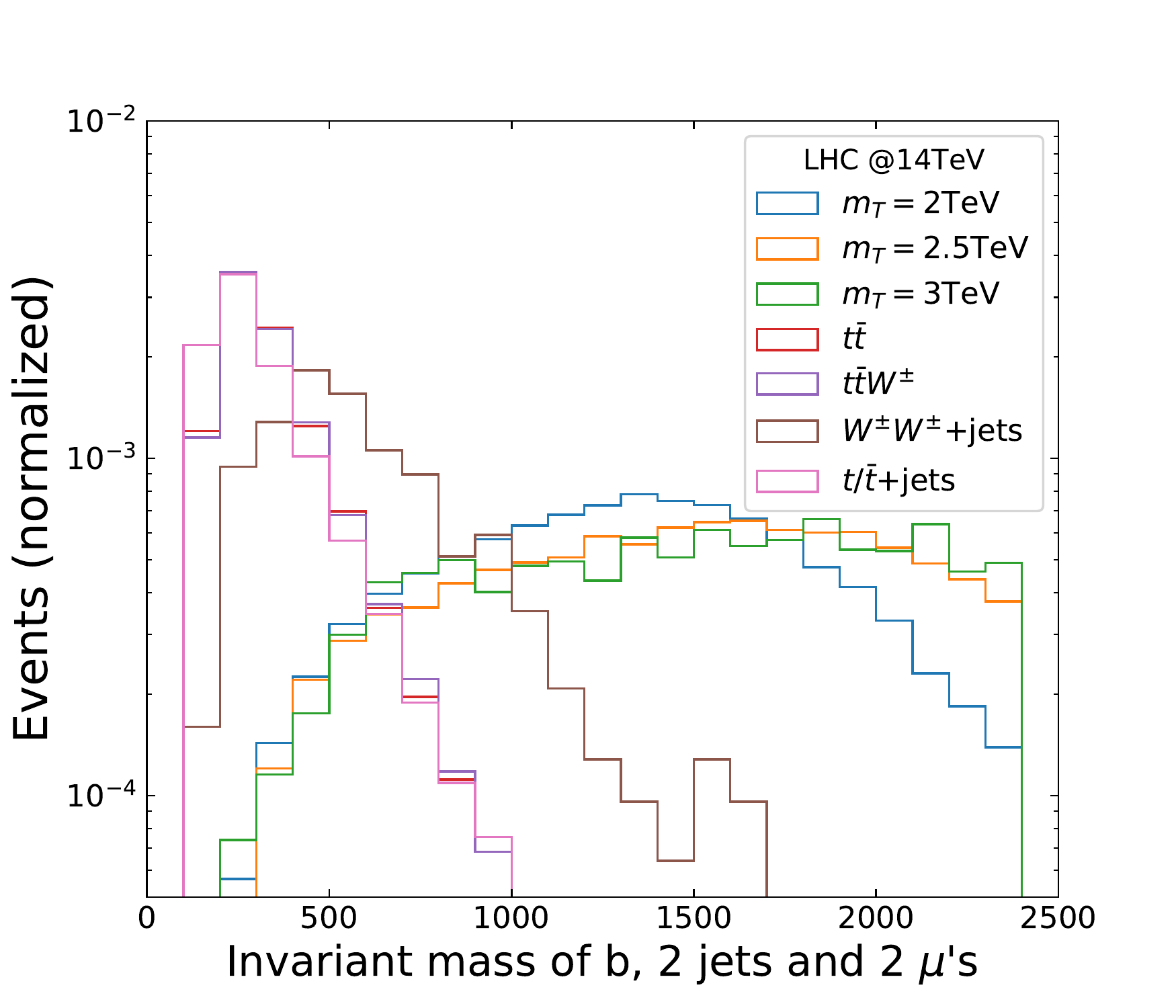}}
  \centerline{(e)}
\end{minipage}
\caption{Kinematic distributions for the signal $pp \to b+\mu^{\pm}\mu^{\pm}$+multijets and the SM backgrounds $pp \to t\bar{t}$, $t/\bar{t}$+jets, $t\bar{t}\,W^\pm$, $WW$+jets at the 14 TeV LHC. The benchmark points are chosen as $m_{N}=50$\,GeV, $V_{\mu N}=1.0$, $m_{T}$=2,\,2.5,\,3\,TeV, $V_{Tb}=0.1$.}
\label{fig:dist}
\end{figure}

Kinematical distributions for signal and SM backgrounds at the 14\,TeV LHC are shown in \figurename~\ref{fig:dist}. Note that for the signal, we present three different benchmark points as $m_{T}=$2, 2.5 and 3\,TeV. \figurename~\ref{fig:dist}(a) presents the product of charges of final two muons, from which we can find that $t\bar{t}$, $t\bar{t}W^{\pm}$ and $t/\bar{t}$+jets events tend to have opposite-sign dimuon. \figurename~\ref{fig:dist}(b) shows the distributions of missing transverse energy in which the curve of signal extends further than those of the backgrounds in the large range of $\slashed{E}_{T}$. Due to the large mass of the VLT, b quarks from $T$ decay are highly energetic and hence by parton showering the neutrinos from the b quarks constitute large $\slashed{E}_{T}$ as shown. From the curves of three benchmark points for our signal, we can find that a larger VLT mass will be reflected in the more flat distribution of $\slashed{E}_{T}$. Distributions of relative distance between final dimuon are displayed in \figurename~\ref{fig:dist}(c), where for the signal events $\Delta R_{\mu\mu}$ is smaller than that for the backgrounds, since the dimuon comes from the same parent particle $T$ in the former case while final muons come from different parent particles in the latter case. In \figurename~\ref{fig:dist}(d) we present distributions for rapidity of the leading jet (non-b-tagged). In the single VLT production, the jet from splitting of a valence quark with $W$ boson emission is always of strong forward nature which can be seen in the rapidity distributions. But cutflows with cuts on rapidity of the forward jet show that it is not effective if other cuts, such as ones on $\slashed{E}_{T}$ and $\Delta R_{\mu\mu}$, are first applied. Thus in the following cuts we do not include this one. \figurename~\ref{fig:dist}(e) displays distributions for transverse momentum of leading b jet, which can be used to well separate signal and backgrounds because b jet from VLT with a mass of 2\,TeV tends to be much harder than that from the SM top quark in background events. We can also find that the VLT with a larger mass leads to a longer tail in the distribution of $p_{T}$ of the leading b jet. Finally we reconstruct the parent VLT mass by $m_{b2\mu2j}$ clustering the leading b jet, the dimuon and two soft jets, the distributions of which in \figurename~\ref{fig:dist}(f) show that more signal events distribute around the range of 500\,GeV$<m_{b2\mu2j}<$2\,TeV, while the background distributions tend to center around the range of a much smaller $m_{b2\mu2j}$.

According to the above distributions and analysis, the following cuts are applied that can well distinguish signal from the SM backgrounds:
\begin{itemize}
\item Cut 1: Two muons of same sign are required and each of them should satisfy $p_{T}(\mu)>10$\,GeV and $|\eta(\mu)|<2.8$.
\item Cut 2: At least 4 jets in the final states are required with $p_{T}(j)>15$\,GeV and $|\eta(j)|<3.0$.
\item Cut 3: We require a large missing transverse energy as $\slashed{E}_{T}>160$\,GeV.
\item Cut 4: Relative distances are required for the dimuon separation as $0.4<\Delta R_{\mu\mu}<1.0$, for jets separation as $\Delta R_{jj}>0.4$ and for jet-muon separation as $\Delta R_{\mu j}>0.4$.
\item Cut 5: At least one of the final jets is required to be a b-tagged one, which also should have $p_{T}>210$\,GeV.
\item Cut 6: The invariant mass $m_{b2\mu2j}>1200$\,GeV is required.
\end{itemize}
\begin{table}
\centering
\begin{tabular}{l|c|c|c|c|c}
\hline\hline
  & $t\bar{t}$ & $t/\bar{t}$+jets & $t\bar{t}\,W^{\pm}$ & $WW+$jets & signal \\ \hline
Cut 1: Same-sign dimuon & 16.3 & 1.46 & $2.84\times10^{-2}$ & $1.76\times10^{-3}$ & 17.6  \\ \hline
Cut 2: No.(jets)$\geqslant4$ & 10.9 & 0.451 & $2.12\times10^{-2}$ & $5.90\times10^{-4}$ & 8.33  \\ \hline
Cut 3: $\slashed{E}_{T}>160$\,GeV  & 0.191 & $1.22\times10^{-3}$ & $2.15\times10^{-3}$ & $1.20\times10^{-4}$ & 1.59  \\ \hline
Cut 4: On relative distances\, & $2.74\times10^{-2}$ & $2.72\times10^{-4}$ & $3.96\times10^{-4}$ & $5.41\times10^{-6}$ & $1.10$  \\ \hline
Cut 5: No.(b)$\geqslant1$ \& $p_{T}>210$\,GeV & $2.70\times10^{-3}$ & $1.95\times10^{-5}$ & $4.87\times10^{-5}$ & $2.62\times10^{-7}$ & $0.528$  \\ \hline
Cut 6: $m_{b2\mu2j}>1200$\,GeV & $1.74\times10^{-5}$ & 0 & $3.22\times10^{-6}$ & $0$ & $0.279$ \\
\hline\hline
\end{tabular}
\caption{Cutflow of cross sections for signal process $pp \to b+\mu^{\pm}\mu^{\pm}$+multijets and the SM background processes $pp \to t\bar{t}$, $t/\bar{t}$+jets, $t\bar{t}\,W^\pm$, $WW$+jets at the 14 TeV LHC. The benchmark point is $m_{N}=50$\,GeV, $V_{\mu N}=1.0$, $m_{T}=2$\,TeV, $V_{Tb}=0.1$. Cross sections are shown in unit of pb.}
\label{tab:cutflow}
\end{table}
In \tablename~\ref{tab:cutflow} we present the cutflow of cross sections for both signal and backgrounds with the above cuts applied, from which we can see that the dominant background is the SM top pair production. With the first two cuts on numbers of final same-sign muons and jets, the effective cross sections of backgrounds can be suppressed to the same order as that of the signal. Requirements on $\slashed{E}_{T}$ and relative distances can then further reduce backgrounds to percent level compared to the signal. Final two cuts on number of b jets and the reconstructed mass can remove backgrounds $t/\bar{t}$+jets and $WW+$jets while the $t\bar{t}$ and $t\bar{t}\,W^{\pm}$ events are left at a negligible level (about 4 orders smaller than the signal). With these cuts for event selection, we can expect a promising search for the VLT new decay channel through its single production.

\begin{figure}[t]
\centering
\begin{minipage}{0.48\linewidth}
  \centerline{\includegraphics[scale=0.65]{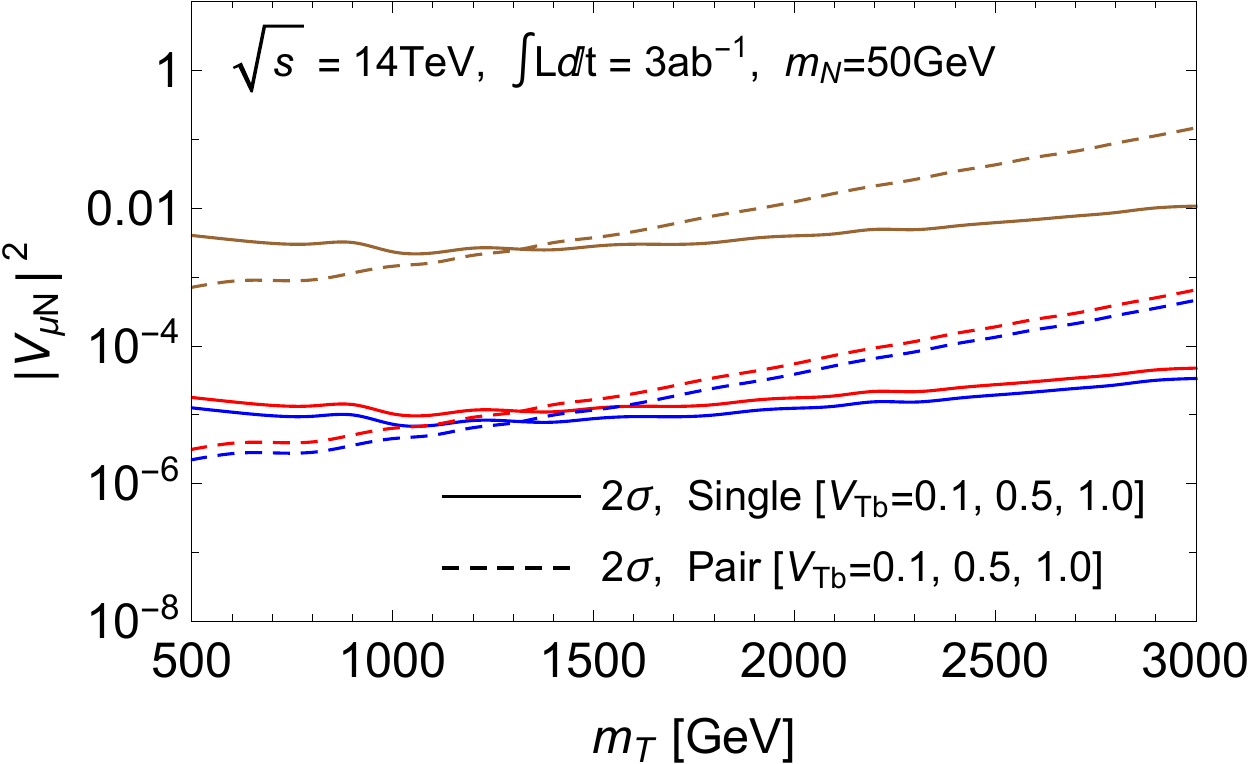}}
  \centerline{(a)}
\end{minipage}
\hfill
\begin{minipage}{0.48\linewidth}
  \centerline{\includegraphics[scale=0.65]{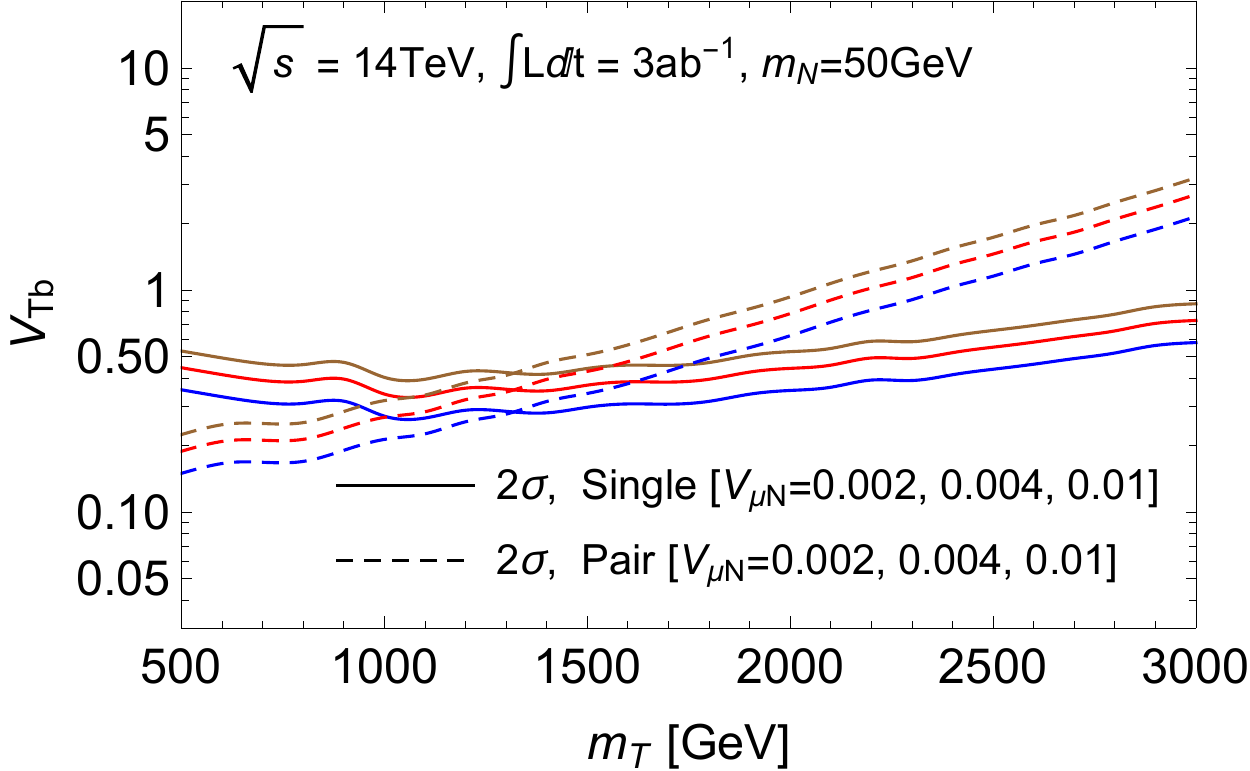}}
  \centerline{(b)}
\end{minipage}
\\[12pt]
\begin{minipage}{0.48\linewidth}
  \centerline{\includegraphics[scale=0.7]{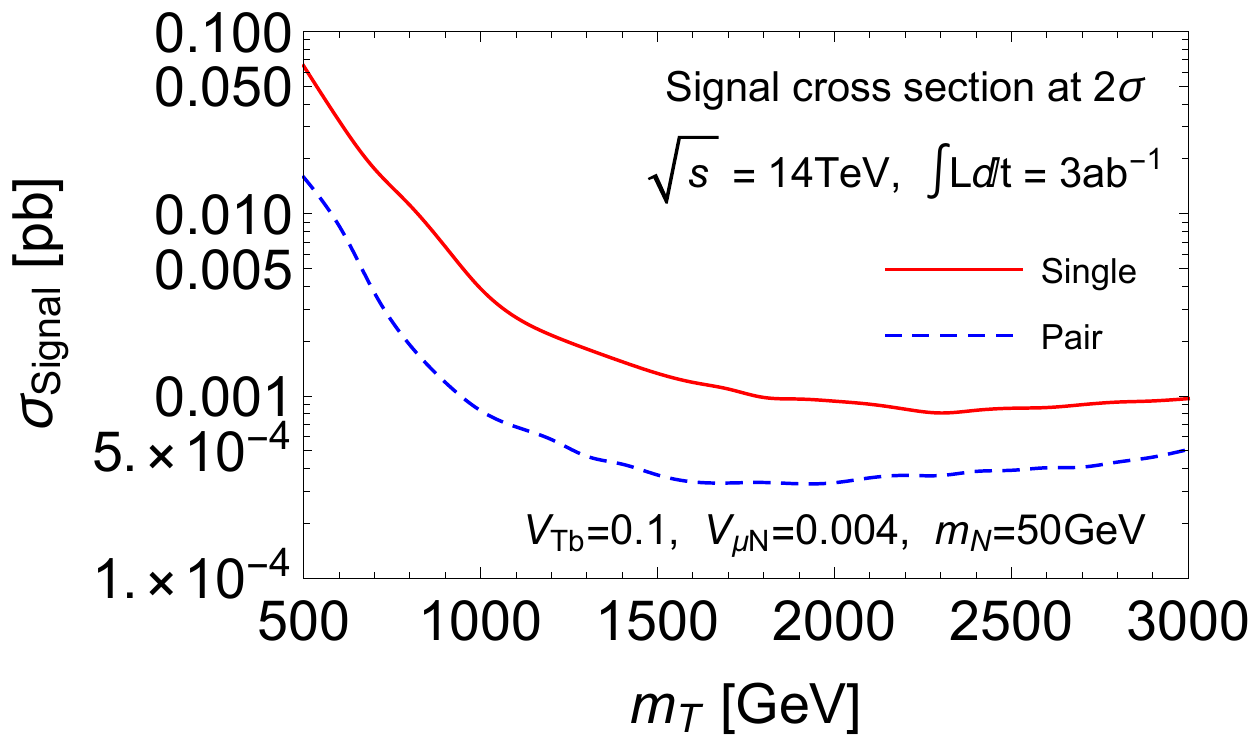}}
  \centerline{(c)}
\end{minipage}
\caption{$2\sigma$ exclusion limits on the VLT single production (solid lines) followed by the new decay channel $pp\to T/\bar{T}+\text{jets}\to b/\bar{b}+\mu^{\pm}\mu^{\pm}+\text{jets}$ at the 14 TeV LHC, with integrated luminosity of $3\,\text{ab}^{-1}$. Systematic uncertainty $\beta$ is taken as 5\%. (a) is plotted on the plane of $|V_{\mu N}|^{2}$ versus $m_{T}$ and (b) on the plane of $V_{Tb}$ versus $m_{T}$. In each figure we also include the results from VLT pair production (dashed lines)~\cite{Zhou:2020ovl} as comparison.}
\label{fig:exclu}
\end{figure}

To calculate the statistical significance, we use the formula $\alpha=S/\sqrt{B+(\beta B)^{2}}$ where $\beta$ is the systematic error and $S$\,($B$) is the number of signal (background) events with the above cuts applied. In \figurename~\ref{fig:exclu} we present the $2\sigma$ exclusion limits on the VLT single production (solid lines) followed by the new decay channel $pp\to T/\bar{T}+\text{jets}\to b/\bar{b}+\mu^{\pm}\mu^{\pm}+\text{jets}$ at the 14 TeV LHC with integrated luminosity of $3\,\text{ab}^{-1}$. In the present case, systematic uncertainty mainly comes from the background with misidentified leptons and is taken as 5\%. In each figure we also include the results from VLT pair production (dashed lines)~\cite{Zhou:2020ovl} as comparison.

\figurename~\ref{fig:exclu}(a) is shown on the plane of neutrino mixing $|V_{\mu N}|^{2}$ versus the VLT mass where three solid lines from top to bottom correspond to cases of $V_{Tb}=0.1$, 0.5, 1.0. In the given VLT mass region, the mixing between $\mu$-flavor and heavy Majorana neutrino $|V_{\mu N}|^{2}$ can be probed to orders of $10^{-6}\sim10^{-5}$ with $V_{Tb}=0.5$ or 1.0, in which the best point for $V_{Tb}=1.0$ can be reached down to $6.7\times10^{-6}$ at $m_{T}\sim1.05$\,TeV. Note that experiments of colliders including the LHC and LEP are able to produce large amount of $W$ bosons and can easily search for the same-sign dilepton events.  Current bounds are given around $|V_{\mu N}|^{2}\sim10^{-5}$ in the range of $10\,\text{GeV}\lesssim m_{N}\lesssim m_{W}$ from the DELPHI Collaboration~\cite{Abreu:1996pa}, as well as the searches at the LHC for same-sign dilepton~\cite{Sirunyan:2018xiv} and trilepton events~\cite{Sirunyan:2018mtv}, which, as seen from \figurename~\ref{fig:exclu}(a), can be well improved in our case for a wide range of $m_{T}$ from 800 to 2000 GeV. In \figurename~\ref{fig:exclu}(b), the contours are displayed on the plane of VLT-SM coupling $V_{Tb}$ versus $m_{T}$ for cases of $V_{\mu N}=0.002,\,0.004,\,0.01$ corresponding to solid lines from top to bottom. In the given mass region, $V_{Tb}$ can be excluded at $2\sigma$ down to $0.26\sim0.39$ for the above three settings of $V_{\mu N}$ with the best point at $m_{T}\sim1.05$\,TeV. We can also find from \figurename~\ref{fig:exclu}(a) and \figurename~\ref{fig:exclu}(b) that, compared with the results of VLT pair production~\cite{Zhou:2020ovl}(dashed lines in each figure), the sensitivity of the VLT single production surpasses that of pair production for the VLT mass larger than 1.3 TeV. $2\sigma$ exclusion bound (solid line) on the cross sections of our signal is also presented in \figurename~\ref{fig:exclu}(c) assuming $V_{\mu N}=0.004$, $m_{N}=50$\,GeV and $V_{Tb}=0.1$, as a comparison with that for the VLT pair production (dashed line) in the given VLT mass region. Note that \figurename~\ref{fig:exclu} are obtained with a kinematical accessible $m_{N}$ as 50 GeV, the results of which can be improved further for a less massive heavy Majorana neutrino since the new VLT decay branching ratio will increase accordingly (\figurename~\ref{fig:fdandbr}(b)).

Finally we comment on the pileup effects in our discussion, which, although need proper removal techniques~\cite{Cacciari:2007fd,Krohn:2013lba,Berta:2014eza} for a fully realistic simulation, have limited effects on our results since the event selection is based on hard same-sign dileptons.

\section{Conclusion}

We study in this paper the search for the new decay mode of a vector-like top partner mediated by the heavy Majorana neutrino ($T\to b\,\ell^{+}\ell^{+}jj$) in a model-independent scenario that includes a singlet VLT into the low-energy Type-I seesaw, through the VLT single production at the 14 TeV LHC with integrated luminosity of $3\,\text{ab}^{-1}$. A pair of same-sign muons and large missing $E_{T}$ are proposed as signatures in the search strategy. Detector-level simulation shows that with a kinematically accessible $m_{N}$, $2\sigma$ exclusion limit can be obtained for the mixing between $\mu$-flavor and the heavy Majorana neutrino as $|V_{\mu N}|^{2}>6.7\times10^{-6}$ with $V_{Tb}\sim1.05$ and a TeV scale $m_{T}$. For the coupling between the singlet VLT and SM $b$ quark $V_{Tb}$, upper limits can be reached to $0.26\sim0.39$ at $2\sigma$ with $V_{\mu N}=0.01\sim0.004$ and $m_{T}\sim1.05$\,TeV. In the VLT mass range larger than 1.3 TeV, the sensitivity of a single production search is better than that of its pair production. Conclusions can then be drawn that, with a kinematically accessible heavy Majorana neutrino, we can expect a promising result to search at the LHC for the new decay of a singlet VLT mediated by the heavy Majorana neutrino through the VLT single production.

\section{Acknowledgments}
This work is supported by the National Natural Science Foundation of China (NNSFC) under grants No.\,11847208 and No.\,11705093, as well as the Jiangsu Planned Projects for Postdoctoral Research Funds under grant No.\,2019K197.

\end{document}